\documentclass[onecolumn,
preprint,
preprintnumbers,
nofootinbib,
amsmath,amssymb,
aps,
prd,
]{revtex4-2}
\usepackage{graphicx}
\usepackage{bm}
\usepackage{hyperref}

\usepackage{multirow}

\DeclareUnicodeCharacter{2212}{\textendash}
\def\be{\begin{equation}}
\def\ee{\end{equation}}
\def\bea{\begin{eqnarray}}
\def\eea{\end{eqnarray}}

\usepackage{makecell}

\newcommand{\Mpl}{M_{\mathrm{pl}}}

\newcommand{\nn}{\nonumber}

\usepackage{xcolor}

\newcommand{\beq}{\begin{equation}}
\newcommand{\eeq}{\end{equation}}
\newcommand{\beqa}{\begin{eqnarray}}
\newcommand{\eeqa}{\end{eqnarray}}

\makeatletter
\newcommand{\Biggg}{\bBigg@{3.5}}
\makeatother

\begin{document}
\count\footins=1000
\interfootnotelinepenalty=0

\begin{flushright}
TUM-HEP-1602/26\\
\vspace*{1cm}
\end{flushright}

\title{The End of the First Act: Spectral Running, Interacting Dark Radiation, and the Hubble Tension in Light of ACT DR6 Data}
\author{Mathias Garny$^1$}\email{mathias.garny@tum.de}
\author{Florian Niedermann${}^2$}\email{florian.niedermann@su.se}
\author{Martin S. Sloth${}^3$}\email{sloth@sdu.dk}
\affiliation{%
\small ${}^1$Physik Department T31, School of Natural Sciences, Technische Universit\"at M\"unchen \\
James-Franck-Stra\ss e 1, D-85748 Garching, Germany\\
${}^2$Nordita, KTH Royal Institute of Technology and Stockholm University\\
Hannes Alfv\'ens v\"ag 12, SE-106 91 Stockholm, Sweden
\\
${}^3$ Universe-Origins, University of Southern Denmark, Campusvej 55, 5230 Odense M, Denmark}%

\begin{abstract}
We point out that constraints on $\Delta N_\mathrm{eff}$ reported by the ACT collaboration in their DR6 data release are surprisingly sensitive to the assumptions made about the initial power spectrum from inflation. The ACT collaboration reports no evidence of new light degrees of freedom alongside a low value of the expansion rate, thus confirming the Hubble tension. However, as we show here, when considering self-interacting dark radiation and including running, $\alpha_s$, and running of the running, $\beta_s$, of the spectral index $n_s$, the picture changes significantly. Confronting this extended model with Planck, ACT DR6, DESI DR2, and uncalibrated Pantheon+ data, we find the significantly relaxed bound $\Delta N_\text{eff}< 0.58$ at 95$\%$ CL, together with a $2.9 \sigma$ ($2.6 \sigma$) preference for $\alpha_s>0$ ($\beta_s>0$), while the Hubble tension is reduced to $2.2 \sigma$ with only three more parameters compared to $\Lambda$CDM. If the dark radiation fluid is initially coupled to dark matter, and undergoes dark radiation-matter decoupling (DRMD) around matter-radiation equality, predicting dark acoustic oscillations with drag horizon $r_{d,\mathrm{DAO}} \approx 60 \,\mathrm{Mpc}/h$, the bound is further relaxed to $\Delta N_\text{eff}< 0.68$ at 95$\%$ CL, reducing the Hubble tension below $2\sigma$. We also discuss how $\alpha_s$ and $\beta_s$ could naturally appear in inflationary scenarios, possibly connected to the end of a first act of inflation. In this case dark radiation is mostly probed by scales covered by Planck and DESI, while smaller scales carry information on inflationary dynamics. 
\end{abstract}

\maketitle

\newpage

\section{Introduction}
The Atacama Cosmology Telescope (ACT) has recently released some of the most precise measurements ever made of the temperature anisotropies in the Cosmic Microwave Background (CMB) on small angular scales and of its polarization \cite{AtacamaCosmologyTelescope:2025vnj,AtacamaCosmologyTelescope:2025blo}. These measurements provide strong constraints on the number of effective extra relativistic degrees of freedom $\Delta N_\mathrm{eff}$ during the epoch leading up to recombination \cite{AtacamaCosmologyTelescope:2025nti} (see also \cite{SPT-3G:2025bzu,Beringue:2025bur,Ferreira:2025lrd,Poulin:2025nfb,Tristram:2025you,Escudero:2026mgw,Cvetko:2025kda,Goldstein:2026iuu} for further analysis and discussions of these results and of their consequences). These results  typically assume free-streaming radiation and are consistent with the number of relativistic degrees of freedom expected from the Standard Model of particle physics within the $\Lambda$ Cold Dark Matter ($\Lambda$CDM) model. 

On the other hand, the ACT data reproduce the mild tension with the Baryon Acoustic Oscillation (BAO) data from DESI that was already seen in Planck data~\cite{DESI:2025zgx}, and the Hubble tension, \textit{i.e.}, the discrepancy between the expansion rate today as inferred indirectly from CMB experiments assuming the $\Lambda$CDM model and measured directly using calibrated supernova (SN) data~\cite{Riess:2021jrx} (for a review see~\cite{CosmoVerseNetwork:2025alb}). In addition, the spectral tilt of primordial perturbations measured by ACT has shifted to be bluer, indicating a slight tension also with earlier CMB data~\cite{Ferreira:2025lrd}.

In order to solve the Hubble tension, an injection of extra energy density before recombination is required to reduce the sound horizon and increase the Hubble rate, while staying consistent with BAO data that constrain the product of the two~\cite{Knox:2019rjx}. A simple and natural candidate for such an energy injection is a component of extra dark radiation, as dark radiation would redshift similarly to the background evolution in the period up to recombination. Although \textit{free-streaming} dark radiation has long been excluded as a solution to the Hubble tension because of its anisotropic stress component and the resulting constraints on $\Delta N_\mathrm{eff}$, models of self-interacting dark radiation perform better overall~\cite{Jeong:2013eza,Buen-Abad:2015ova,Buen-Abad:2017gxg,Archidiacono:2020yey,Blinov:2020hmc,Aloni:2021eaq}, provided the dark radiation is dominantly produced after Big Bang nucleosynthesis (BBN)~\cite{Garny:2024ums,Allali:2024cji,Aloni:2023tff,Schoneberg:2022grr}. A conceptually simple and promising model for solving the Hubble tension is the recently proposed model of Dark Radiation-Matter Decoupling (DRMD)~\cite{Garny:2025kqj} within Hot New Early Dark Energy\footnote{The New Early Dark Energy (NEDE) framework for solving the Hubble tension by an energy injection from a fast-triggered phase transition between BBN and recombination was first proposed in \cite{Niedermann:2019olb,Niedermann:2020dwg}. Different models within the NEDE framework are possible, as also discussed in these first papers. The NEDE model, studied extensively in~\cite{Niedermann:2019olb,Niedermann:2020dwg}, where the trigger is an ultralight scalar field is now called Cold NEDE to discriminate it from Hot NEDE, where the trigger is the dark-sector temperature. But for example also a hybrid type model with a slow roll-over is possible~\cite{Niedermann:2020dwg}. For an overview of the framework and other works on Cold NEDE see~\cite{Cruz:2022oqk,Niedermann:2023ssr,Cruz:2023lmn,Chatrchyan:2024xjj,Cruz:2023cxy}.}
(Hot NEDE~\cite{Niedermann:2021vgd, Niedermann:2021ijp, Garny:2024ums}; see also \cite{Buen-Abad:2024tlb,Buen-Abad:2025bgd,Barron:2026nks} for models with a similar cosmological evolution but different microphysical realization based on atomic dark matter~\cite{Cyr-Racine:2012tfp}). In the regime where it addresses the Hubble tension most efficiently, this model also resolves the BAO-CMB tension and predicts the existence of dark acoustic oscillations (DAO) with a drag horizon $r_{d,\mathrm{DAO}} \approx 60 \,\mathrm{Mpc}/h$ \cite{Garny:2025kqj,Garny:2025szk,Garny:2026ish}, which imprint themselves as additional oscillatory features in the matter power spectrum and CMB. Crucially, the DAO feature is localized at intermediate CMB scales, best probed by Planck 2018 data, which thus drives the evidence for it reported in \cite{Garny:2026ish}. On the other hand, while ACT is not expected to have much direct constraining power on this DAO feature itself, it can however constrain the model indirectly through the impact of $\Delta N_\mathrm{eff}$ on small scales (see for example the recent analysis~\cite{Cvetko:2025kda} for a model with similar phenomenology).

Specifically, the ACT collaboration reports~\cite{AtacamaCosmologyTelescope:2025nti} $N_\mathrm{eff}= 2.86\pm0.13$, which corresponds to $\Delta N_\mathrm{eff} = N_\mathrm{eff} - N_\mathrm{eff}^{\rm SM} = -0.18 \pm 0.13$, assuming $N_\mathrm{eff}^{\rm SM} = 3.044$. However, resolving the Hubble tension requires $\Delta N_\mathrm{eff} \gtrsim 0.3$.
Superficially, these new strong constraints therefore seem to pose a problem for an important class of models that can otherwise resolve both the Hubble and the BAO-CMB tensions~\cite{Cvetko:2025kda}. There are, however, some points that should prompt caution before jumping to such conclusions. ACT data indicate a bluer spectrum and a preference for a positive spectral running, which indeed is in conflict with the stronger radiation-induced damping in the tail of the acoustic oscillations from extra relativistic degrees of freedom, and drives the strong ACT constraints on $N_\mathrm{eff}$. However, instead of lowering  $N_\mathrm{eff}$ to gain more power on small scales and remain consistent with ACT data, there are other possible explanations for the increased power ACT sees on small scales.

Since this regime is foreground-dominated, the extent to which it can be used for extracting  cosmological information relies on accurate foreground removal. While we take ACT DR6 data at face value in this work, relying on the sophisticated foreground modelling of the ACT collaboration, one may also consider the hypothesis that the extra power in DR6 data is due to a residual foreground effect. In this context, it is interesting to note that the South Pole Telescope (SPT) \cite{SPT-3G:2025bzu} does not find evidence for a bluer spectral index, and their constraints on $N_\mathrm{eff}=3.18^{+0.29}_{-0.33}$ are looser than previous constraints from Planck $N_\mathrm{eff}=2.86\pm 0.19$. As a side remark, one may note that spectral running has been reported in previous data releases, in that case by SPT and with an opposite sign \cite{Hou:2012xq}, and gone away in their current data release. Together, these facts should serve as a reminder to be cautious before taking the implications of ACT DR6 data for granted.

A second possible explanation is new physics. Additional power on small scales could arise from non-trivial dynamics during inflation leading to a bluer primordial spectrum on small scales. In order to check for such an effect, it is not enough to include running of the spectral tilt, as this is already constrained on large scales by Planck data. One would need to allow for running of running, such that the effect mainly shows up on small scales. Here we are going to investigate this possibility. What happens to the ACT constraints on models which resolve the Hubble tension with a large $N_\text{eff}$, when we allow for running and running of running? As we will show in Sec.~\ref{datasec}, it turns out that within this arguably non-minimal scenario, the combination of current CMB, BAO and uncalibrated SN data, including ACT DR6, allows for a full resolution of the Hubble tension in the DRMD model, and an almost complete resolution in models of (fluid-like) self-interacting dark radiation (SIDR), which is a subset of the DRMD model.

\begin{figure}[t]
  \centering
    \includegraphics[width=0.49\textwidth]{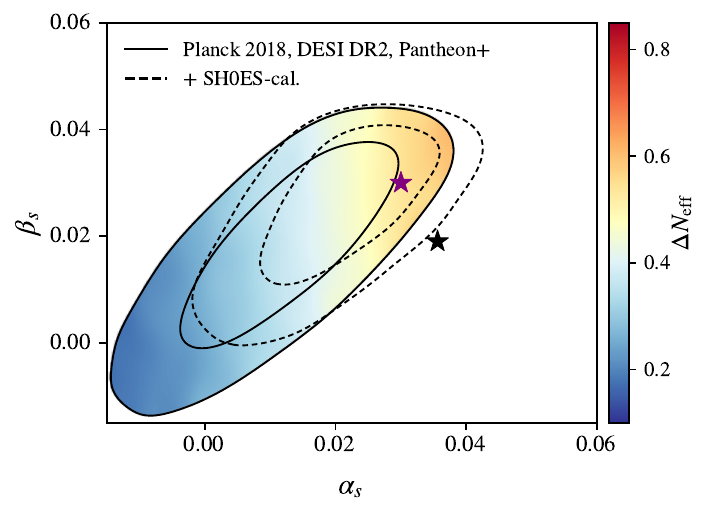}\hfill
        \includegraphics[width=0.49\textwidth]{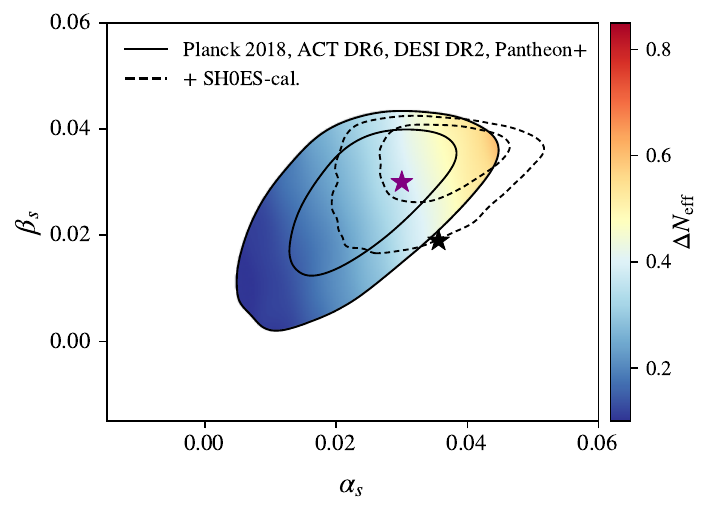}\hfill      
  \caption{Marginalized 2D posteriors for running $\alpha_s$ and running-of-running $\beta_s$ in a model with fluid-like extra radiation (SIDR) from a combination of CMB, DESI DR2 BAO and uncalibrated Pantheon+ SN data (solid lines). CMB data in the left panel correspond to Planck 2018 while the right panel includes also ACT DR6. Dashed contours supplement the analysis by the SH$0$ES calibration of the SN absolute magnitude for comparison. Strong spectral running enables large values of $\Delta N_\mathrm{eff}$ (thus shifting the inferred $H_0$ to larger values compared to $\Lambda$CDM) both for Planck alone and combined with ACT data. The black and purple stars denote the benchmark points~\eqref{eq:benchmark} and~\eqref{eq:benchmark2}, respectively, within the specific inflationary models considered in this work. }
  \label{fig:running_running}
\end{figure}

We adopt the following ansatz for the primordial scalar power spectrum:
\begin{equation}\label{eq:spectral_ansatz}
\mathcal{P}_\mathcal{R}(k)
=
A_s\,
\exp\!\left[
(n_s - 1)\,\ln\!\left(\frac{k}{k_\ast}\right)
+ \frac{1}{2}\,\alpha_s \,\ln^2\!\left(\frac{k}{k_\ast}\right)
+ \frac{1}{6}\,\beta_s \,\ln^3\!\left(\frac{k}{k_\ast}\right)
+ \ldots\right]\,,      
\end{equation}
where $n_s$ is the spectral tilt, $\alpha_s$ its running, and $\beta_s$, the running of the running, defined as
\begin{align}\label{taylorrunning}
n_s-1=\frac{d\ln \mathcal{P}_\mathcal{R}}{d \ln(k)}\Bigg|_{k=k_*} \,, && \alpha_s=\frac{dn_s}{d \ln k}\Bigg|_{k=k_*}\,, &&\beta_s=\frac{d \alpha_s} {d \ln k}\Bigg|_{k=k_*}\,.
\end{align}
The ellipsis stands for higher-order terms in the Taylor expansion and $k_*$ is the pivot scale. In this work, we confront the DRMD and SIDR models with ACT DR6 data along with Planck 2018, DESI DR2 BAO and uncalibrated Pantheon+ SN, allowing for running and running of the running. As a main result of this work, we find that these data sets prefer a running spectral index with $\alpha_s \sim \beta_s \approx 0.03$, corresponding to a preference for $\alpha_s>0$ ($\beta_s>0$) at $2.9 \sigma$ ($2.6 \sigma$). As shown in Fig.~\ref{fig:running_running}, in the simple SIDR model, this is accompanied by a sizeable amount of (fluid-like) extra radiation.  This positive correlation between spectral running and $\Delta N_\mathrm{eff}$ is already present in Planck data (left panel), but only becomes a significant preference over $\Lambda$CDM once ACT data are included (right panel). As a consequence, the Hubble tension is reduced to $2.2\sigma$ in the presence of ACT data. These results are further enhanced in the DRMD model, where the bound on $\Delta N_\mathrm{eff}$ is further relaxed, yielding a preferred value of $H_0 = 70.3^{+1.0}_{-1.2}\,\mathrm{km}\,\mathrm{s}^{-1}\,\mathrm{Mpc}^{-1}$ at 68\% CL. This model also predicts a DAO feature with a drag horizon $r_{d,\mathrm{DAO}} \approx 60\,\mathrm{Mpc}/h$, consistent with previous analyses that do not include ACT data~\cite{Garny:2025szk,Garny:2026ish}. 

The apparent preference for running and running of the running in ACT data could by itself be indicative of some residual foreground effect at play, but it could also very naturally follow from inflation. It has long been thought that $60$ $e$-folds of uninterrupted single field slow-roll inflation is unnatural and requires fine-tuning in the inflaton potential, and a more natural solution is that inflation happens in several acts, interrupted by new dynamics and only to continue in a new direction in field space\footnote{Another possibility is, instead of being completely interrupted, inflation continues in a non-perturbative back-reaction regime \cite{Green:2009ds,Watanabe:2009ct,Adshead:2012kp,Notari:2016npn,Gorbar:2021rlt,Figueroa:2023oxc,Iarygina:2023mtj,Figueroa:2024rkr,Jamieson:2025ngu}. In that case, the end of the first act should be thought of as the end of the perturbative regime.} \cite{Silk:1986vc,Polarski:1992dq,Adams:1997de,Lyth:1995ka,Burgess:2005sb,Dvali:2003vv,Freese:2004vs,Easther:2004ir,DAmico:2020euu,DAmico:2021vka,DAmico:2021fhz,DAmico:2026sfz}. At the end of the first act, which is where the observable modes in the CMB exit the horizon, one expects indication of departures from single field slow-roll inflation to start appearing. In single field slow-roll, the running and the running of running is second order and third order in slow-roll, respectively, and therefore expected to be small. Indication of large running and running of running would therefore indicate that the first act of inflation is ending, and inflation occurs in several stages.

Below, in Sec.~\ref{exsec} we give two examples of non-trivial dynamics, which could lead to an interruption of inflation. One is the commonly considered possibility that the inflaton couples to gauge fields and triggers their resonant production, which 
then back-reacts on the inflaton perturbations, leading to a feature in the inflationary power-spectrum. The second example is the indirect effect of heavy fields, which are sub-dominant and do not contribute to the energy density during inflation, but instead appear as an explicit time-dependence of the inflaton potential. If the time-dependence from heavy fields is large enough, this could lead to large running and running of the running in the primordial spectrum. Such a strong time-dependence can come from heavy fields undergoing a period of fast-roll due to an instability\footnote{For other models of inflation with large running see \cite{Kobayashi:2010pz,Takahashi:2013tj,Czerny:2014wua,Das:2022ubr,Cabass:2016ldu,vandeBruck:2016rfv,Fairbairn:2025fko}.}. 

In both of these examples, the sharp increase in the power spectrum on small scales could be accompanied by other interesting observational effects, such as non-Gaussian perturbations~\cite{Abolhasani:2010kn,Lyth:2012yp,Barnaby:2011qe,Barnaby:2011vw,Linde:2012bt,Ferreira:2014zia,Ferreira:2015omg,Arkani-Hamed:2015bza,Lee:2016vti,Chen:2016uwp,Meerburg:2016zdz,Chen:2022vzh,Pajer:2024ckd}, gravitational waves \cite{Garcia-Bellido:2007fiu,Dufaux:2008dn,Barnaby:2011qe,Ferreira:2014zia,Namba:2015gja,Ferreira:2015omg,Garcia-Bellido:2016dkw,Cheng:2018yyr,Ozsoy:2020kat,DAmico:2026sfz}, and primordial black hole production \cite{Garcia-Bellido:1996mdl,Kawasaki:1997ju,Kawasaki:1998vx,Leach:2000ea,Lyth:2011kj,Lyth:2012yp,Linde:2012bt,Garcia-Bellido:2016dkw,Domcke:2017fix,Erfani:2015rqv,Cheng:2015oqa,Garcia-Bellido:2016dkw,Domcke:2017fix,Cheng:2018yyr,Ozsoy:2020kat,Unal:2023srk,He:2025ieo}. We also highlight that, from a theoretical perspective, the DRMD model (which includes the simpler SIDR model as a subset) is based on a microscopic framework featuring a dark sector described by a dark non-Abelian gauge symmetry, which aligns well with the possible origin of spectral running during a first stage of inflation due to gauge-field dynamics, providing the possibility for a coherent theoretical framework based on fundamental principles known to exist in Nature.

The work is organized as follows. In the next section we give examples of simple natural models of inflation with large running in the primordial spectrum. In Sec.~\ref{datasec}, we present results from including ACT DR6 data in the analysis of DRMD and SIDR models, allowing for spectral running. We conclude in Sec.~\ref{concl}.

\section{Examples of natural large running and running of running}\label{exsec}

In a minimal single field slow-roll inflationary scenario, the solution of the causality problem and the horizon problem requires, depending somewhat on the reheating history, around $N=60$ $e$-folds of inflation in total. It has often been argued that $N=60$ $e$-folds of uninterrupted single-field slow-roll inflation is unnatural from a model building point of view, as it requires a fine-tuned inflaton potential. Instead, inflation might have happened in several shorter stages interrupted by additional new dynamics, leading to the possibility of observable departures from the minimal single field inflation predictions in the CMB \cite{Silk:1986vc,Polarski:1992dq,Adams:1997de,Lyth:1995ka,Burgess:2005sb,Dvali:2003vv,Freese:2004vs,Easther:2004ir,DAmico:2020euu,DAmico:2021vka,DAmico:2021fhz,DAmico:2026sfz}.

If the first stage of inflation lasted around $7-10$ $e$-folds, then in the smallest scales of the CMB spectrum, as measured by ACT, we would be probing the end of the first stage of inflation. During this epoch, the potential steepens and the inflaton accelerates and picks up velocity leading to large running and running-of-running at small scales.  In fact, when the running-of-running becomes of the same order as the running itself, the Taylor expansion in (\ref{taylorrunning}) breaks down, signalling the end of the perturbative slow-roll regime.

The evidence we find in ACT for large running and running-of-running when addressing the Hubble tension and the CMB-BAO tension with SIDR and DRMD, could therefore be considered to be evidence of the end of the first act of inflation. Below we consider two examples of how the first act could end and how such a scenario would manifest itself in the primordial spectrum.

\subsection{Gauge field production}

For simplicity we will assume that during the first act of inflation, inflation is driven by an axion-like field $\phi$, which couples to gauge fields with a coupling of the form
\begin{equation}
\mathcal{L}_{\mathrm{int}}=- \frac{\phi}{4 f_\phi} F^a_{\mu\nu} \tilde{F}^{a\mu \nu}~,
\end{equation}
with effective axion decay constant $f_\phi$, gauge field strength $F_{\mu\nu}^a$ and its dual $\tilde F^{a\mu\nu}$.

It is well-known that in the case where the time-dependence of the inflaton field cannot be ignored, the time-dependence of the inflaton field enters into the equations of motion of the gauge fields and triggers a possible resonant production of gauge fields. The exponentially growing occupation number of gauge fields feed back into the inflaton equation of motion at one-loop level and lead to an amplification of inflaton perturbations, or equivalently of the co-moving curvature perturbation $\zeta$\footnote{Note that the induced coupling to the co-moving curvature perturbation, $\mathcal{L}_{\zeta A^2}=-2 \xi \epsilon^{i j k} \zeta A_i^{a \prime} A_{j, k}^a$, is universal and independent of whether the axion-like particle present during inflation, $\phi$, is the actual inflaton or not \cite{Ferreira:2014zia}.} \cite{Barnaby:2010vf,Meerburg:2012id,Ferreira:2014zia,Namba:2015gja,Ferreira:2015omg,Caravano:2022epk,Domcke:2023tnn,Iarygina:2023mtj,Ishiwata:2025wmo}.

In the Abelian $U(1)$ case\footnote{Inspired by the Hot NEDE and the DRMD model, one may want to consider a non-Abelian $SU(N)$, in which case there is an extra factor of $N^2-1$ multiplying $\kappa_\xi$ \cite{Ferreira:2015omg}.} it was found that the power spectrum ${\cal P}_\zeta$ is modified compared to the case without particle production, denoted by ${\mathcal{P}_\zeta^{(0)}}$,  as~\cite{Barnaby:2011vw,Barnaby:2011qe}
\begin{equation}\label{P_zeta}
    \mathcal{P}_\zeta = \mathcal{P}^{(0)}_\zeta(1+\kappa_\xi)\,,
\end{equation}
where
\begin{equation}\label{eq:kappa}
    \kappa_\xi = \gamma_s \frac{\mathcal{P}^{(0)}_\zeta}{\xi^d}e^{4\pi\xi}
\end{equation}
describes the effect of the resonant particle production with $\xi =\dot\phi/(2f_\phi \,H)$ being the rescaled field velocity. Here, we use $\gamma_s \approx 3 \times 10^{-5}$ and $d=5.4$~\cite{Barnaby:2011qe}. 

Now assuming $\kappa_\xi \ll 1$ at the pivot scale $k_\ast$, we have
\begin{equation}
    n_s-1\equiv \frac{d\ln\mathcal{P}_\zeta}{d\ln k} = n_s^{(0)}-1 +\lambda_\xi \, \kappa_\xi\,,
\end{equation}
where we introduced the spectral tilt in the absence of  particle production
\begin{equation}
    n_s^{(0)}-1 \equiv \frac{d\ln\mathcal{P}^{(0)}_\zeta}{d\ln k}\,,
\end{equation}
alongside 
\begin{equation}\label{eq:lambda0}
    \lambda_\xi \equiv \frac{d\ln\kappa_\xi}{d\ln k} = (-d  +4\pi\xi)\,\delta_\xi + n_s^{(0)}-1 \,,
\end{equation}
with $\delta_\xi \equiv d\ln \xi/d\ln k$.
Assuming that 
\begin{align}\label{eq:assumption}
\frac{d \ln\lambda_\xi}{d\ln k} \ll \frac{d\ln\kappa_\xi}{d\ln k} =\lambda_\xi\,,
\end{align}
we obtain  (neglecting the running of $n_s^{(0)}$)
\begin{equation}\label{eq:alpha_beta}
     \alpha_s \equiv \frac{d n_s}{d \ln k} \approx \lambda_\xi^2 \kappa_\xi \,, \qquad  \beta_s \equiv \frac{d \alpha_s}{d \ln k} \approx \lambda_\xi^3 \kappa_\xi\,,
\end{equation}
at the pivot scale, where analogous expressions hold for higher-order running terms. This implies that 
\begin{align}\label{eq:equal_running}
\beta_s= \lambda_\xi \alpha_s=\lambda_\xi^2\,(n_s-n_s^{(0)}) \approx \lambda_\xi^3 \kappa_\xi.
\end{align}
In other words, as $\lambda_\xi \to 1$ the higher-order Taylor coefficients in \eqref{eq:spectral_ansatz} are all of the same order, indicating the breakdown of the Taylor series for $\ln(k/k_*) \gtrsim 1$. Resumming it, we obtain
\begin{equation}\label{eq:spectral_ansatz_resum}
\mathcal{P}_{\mathcal \zeta}(k)
=
A_s
\left(\frac{k}{k_*}\right)^{\,n_s-1-\alpha_s^2/\beta_s}
\exp\!\left[\frac{\alpha_s^3}{\beta_s^2}\left(\left(\frac{k}{k_*}\right)^{\beta_s/\alpha_s}-1\right)\right] \,.    
\end{equation}
This expression is applicable in a $k$-regime where the form \eqref{P_zeta} is valid (although it cannot be trusted for $k> k_b$, where $k_b$ is the mode that exits the horizon at the moment of perturbative breakdown~\cite{Ferreira:2015omg}).
While it would be interesting to explore the phenomenology of this resummed form, for the purpose of this work, we will assume $\lambda_\xi = \beta_s/\alpha_s\lesssim 1$ and $\ln(k/k_*) \lesssim 1$, implying that the Taylor series can be truncated as in \eqref{eq:spectral_ansatz} (keeping us also within the perturbative regime). In Fig.~\ref{fig:residuals}, we show explicitly that the truncation error for our best-fit cosmologies (orange-shaded band) remains small in the regime that is probed by the data sets used in this work, although more precise small-scale CMB or LSS data will require a more complete description on scales indicated by the gray vertical bands. The truncation error on large-scales, on the other hand, is naturally hidden within the cosmic variance uncertainties. In the next step, we will show how the spectral running pattern in \eqref{eq:alpha_beta} can be achieved in a realistic model.  

In axion monodromy models, inflation is of the large-field type~\cite{Silverstein:2008sg,McAllister:2008hb,Kaloper:2008fb}. If for simplicity we assume that the first stage of inflation is driven by a model of chaotic inflation with ~\cite{Linde:1983gd} 
\begin{equation}
    V(\phi)=\lambda \, \Mpl^4\left(\frac{\phi}{ \Mpl}\right)^2\,,
\end{equation}
then $N$ $e$-folds before the naive end of inflation, the slow-roll parameters are
\begin{align}
\epsilon
    &\equiv \frac{ \Mpl^2}{2}\left(\frac{V'}{V}\right)^2
      = \frac{2  \Mpl^2}{\phi^2}
      = \frac{1}{2N},
\\
\eta
    &\equiv  \Mpl^2 \frac{V''}{V}
      = \frac{2  \Mpl^2}{\phi^2}
      = \frac{1}{2N}.
\end{align}
With the usual expression for the spectral tilt, $n_s^{(0)}-1= -6\epsilon +2\eta = -2 /N$, we obtain
\begin{align}
n_s=1-\frac{2}{N} + \lambda_\xi \kappa_\xi\,.
\end{align}
With $3 H \dot \phi=-V'$ we can rewrite $\xi =\dot\phi/(2f_\phi \,H)$ as
\begin{align}\label{eq:xi}
\xi = \frac{ \Mpl}{f_\phi} \frac{1}{2\sqrt{N}}\,,
\end{align}
which in turn allows us to evaluate \eqref{eq:lambda0} 
\begin{equation}\label{eq:lambda}
    \lambda_\xi\approx (4\pi-d/\xi)\,\frac{\, \Mpl}{4f_\phi}\frac{1}{N^{3/2}} - \frac{2}{N}\,,
\end{equation}
where we used $d\epsilon/d\ln k = -d\epsilon/dN = 1/(2N^2)$. We can now check the validity of the assumption in \eqref{eq:assumption}. From \eqref{eq:xi} and \eqref{eq:lambda}, we obtain the sufficient condition
\begin{align}\label{eq:cond_N_f}
\frac{\sqrt{N}f_\phi}{\pi \Mpl } \ll1\,.
\end{align}
To provide a numerical example, Eqs~\eqref{eq:kappa}, \eqref{eq:equal_running}, \eqref{eq:xi}, and \eqref{eq:lambda} predict 
\begin{align}\label{eq:benchmark}
n_s=0.97\,, \quad \alpha_s= 0.0356\,, \quad \text{and} \quad \beta_s = 0.019,
\end{align}
for the parameter choice $N_*=20.7$, and $f_\phi/ \Mpl= 0.04$ (corresponding to $\epsilon=0.024$, $\lambda_\xi= 0.53$, $\kappa_\xi = 0.12$, and $\xi_* = 2.5$), where the asterisk denotes evaluation at the pivot scale.  These values are compatible with \eqref{eq:cond_N_f} and fall roughly within the $95\%$ confidence intervals we derive in our combined analysis with ACT data (see black star in Fig.~\ref{fig:running_running}). So in this very simple toy model, ACT data non-trivially constrains the axion decay constant to be sub-Planckian while pushing the Taylor expansion of the primordial spectrum to the edge of its range of validity. 

In this toy model perturbations will become order one and inflation will end, before $\xi$ reaches a value of $\mathcal{O}(10)$. The value for which the one-loop gauge-field-induced perturbations become of order the tree-level inflaton perturbations, can be estimated using the condition $\kappa_\xi \approx 1$, which due to \eqref{eq:kappa} yields $\xi \approx 3$ (which applies more universally for rolling axions during inflation \cite{Ferreira:2014zia}). This is on the edge of the perturbative breakdown, which has been estimated to happen slightly later for $ \xi\gtrsim \xi_b = 3.5$ \cite{Ferreira:2015omg} (see also \cite{Ishiwata:2025wmo}), which corresponds to $N_b\approx 10.6$. Thus the first act of inflation in this model will last only $\Delta N =  N_*-N_b \approx 10.1 $ $e$-folds from when modes leave the horizon at the pivot scale. Here, we take $k_* = 0.05/ \mathrm{Mpc}$ in agreement with the \texttt{CLASS} convention, corresponding to $\ell_* \approx 650$ and amounting to a total duration of $\ln(650)+10.1 \approx 16.6$ $e$-folds.\footnote{The  pivot scale corresponds to the choice of expansion point in \eqref{eq:spectral_ansatz}. It is \textit{a priori} arbitrary and a different choice will lead to different values of $\alpha_s$ and $\beta_s$, which should amount to the same total length of the first stage of inflation for the same underlying model parameters. However, its value in general affects the convergence of the truncated Taylor series, and not every choice might be equally convenient.} This, however, is enough to cover the modes observed in the CMB. We stress again that in this setup, there has to be another stage of inflation to ensure that the total number of $e$-folds is large enough to solve the horizon problem. 

Finally, the gauge-field instability considered here is also expected to generate non-Gaussianities. While a systematic analysis is left for future work, we note that the benchmark value $\xi_*=2.5$ is compatible with existing bounds in the literature~\cite{Meerburg:2012id,Planck:2019kim}.

\subsection{Running by proxy}

In single field slow-roll inflation, it is assumed that the inflaton is light compared to the Hubble scale during inflation. However,
during inflation sub-dominant heavy degrees of freedom can manifest themselves indirectly in many different ways \cite{Achucarro:2010jv,Achucarro:2012sm,Cespedes:2012hu,Achucarro:2012yr,Cespedes:2013rda,Pi:2012gf,Achucarro:2012fd,Konieczka:2014zja,Gao:2015aba,Arkani-Hamed:2015bza,Lee:2016vti,Chen:2016uwp,Meerburg:2016zdz,Chen:2022vzh,Pajer:2024ckd}.

One possibility is that, as a field becomes heavy and falls out of slow-roll, it starts to fast-roll \cite{Linde:2001ae}, and if it couples to the inflaton field, it leads to a strong time-dependence in the effective inflaton potential \cite{Jain:2015jpa}. The explicit time-dependence of the inflaton potential will induce a running in the spectral index, called {\it running by proxy} \cite{Sloth:2014sga,Jain:2015jpa}, and generically also running of the running. In the following we discuss two examples, one in which the heavy fields affect the inflation, and one in which a similar effect occurs for a curvaton field \cite{Enqvist:2001zp,Lyth:2001nq,Moroi:2001ct}.

\subsubsection{Inflaton with proxy running}

We can effectively capture the dynamics of such heavy fields by considering only their indirect effect, such that they appear just as an explicit time-dependence in the inflaton potential, $V(\phi,N)$, where $\phi$ is the inflaton and $N\approx Ht$ is the number of $e$-folds. In this case, using the standard relation $k=aH$ at horizon crossing and the slow-roll equations for $\phi$, we can write $d/d\ln k = -(V'/V)\partial_\phi+\partial_N$, where prime denotes the partial derivative with respect to $\phi$, $\partial_\phi$, and a $N$-subscript denotes partial derivatives with respect to $N$, $\partial_N$, below. 

To  leading order in slow-roll, we then obtain \cite{Jain:2015jpa}
\bea
n_s-1&=& -6\epsilon +2\eta +3 \frac{V_N}{V}-2\frac{V_N'}{V'}-2\Gamma_0\frac{\Mpl^2\,V''_N}{V}\,,\\
\alpha_s &=& 4\frac{\Mpl^2\,V''_N}{V}+3 \frac{V_{NN}}{V}-2\frac{V_{NN}'}{V'}-2\Gamma_0\frac{\Mpl^2\,V''_{NN}}{V}\,,\\
\beta_s&=& 6\frac{\Mpl^2\,V''_{NN}}{V}+3 \frac{V_{NNN}}{V}-2\frac{V_{NNN}'}{V'}-2\Gamma_0\frac{\Mpl^2\,V''_{NNN}}{V}\,,
\eea
where $\Gamma_0 = 2 - \ln(2) - \gamma_E \approx 0.73$ and in the slow-roll expansion we assumed that $V_N/V$, $V'_N/V \lesssim \mathcal{O}(1/100)$ in order for the spectral index to remain close to flat, counting as same order as the slow-roll parameters $\epsilon$ and $\eta$,  while extra derivatives of $N$ can be unsuppressed in the slow-roll counting, when the potential has a strong explicit time-dependence, as we are assuming when running and running of running is of the same order as the spectral tilt.

As a simple example we consider chaotic inflation with a small time-dependent mass
\begin{equation}
    V(\phi,N) = \frac{1}{2}(m_\phi^2 +\Delta m^2(N))\phi^2 \,.
\end{equation}
We assume that the time-dependent mass is small $\Delta m(N) \ll m_\phi$, and since this is a large-field model, we can also assume $\phi/\Mpl \gg 1$. In this model the spectral index then becomes 
\bea
    n_s-1 &=& -6\epsilon+2\eta  +2\frac{\Delta m\Delta m_N}{m_\phi^2} \,,\\
    \alpha_s &=& 2\frac{\Delta m\Delta m_{NN}+\Delta m_N^2}{m_\phi^2}\,,\\
    \beta_s &=& 2\frac{\Delta m_N\Delta m_{NN}+\Delta m \Delta m_{NNN}+2\Delta m_N\Delta m_{NN}}{m_\phi^2}\,.
\eea

Now if the time-dependence of the potential comes from a sub-dominant, but heavy scalar field, $\chi$, in fast-roll, that we model by way of example as being described by a tachyonic instability, then we can write the potential of the heavy field as
\beq
V(\chi) =V_0 -\frac{1}{2}m_\chi^2\chi^2\,,
\eeq
where the solution to lowest order in slow-roll is \cite{Linde:2001ae}
\be\label{solchi}
\chi=\chi_0 e^{F\left(m_\chi / H\right)Ht},
\ee
where $F\left(m_\chi^2 / H^2\right)=\sqrt{9 / 4+m_\chi^2 / H^2} -3/2$.
This means that if we take, as a simple example, $\Delta m^2 = \lambda_\chi \, \chi^2$, we obtain, after integrating out the heavy $\chi$ field, an effective single-field inflaton potential with a time-dependent mass
\be
\Delta m(N) = \lambda_\chi^{1/2}\chi_0 \exp\left(F\left(m_\chi / H\right)N\right)\,,
\ee
assuming that $\lambda_\chi \phi^2 \ll m_\chi^2$. This implies
\be\label{eq:by_proxy}
    n_s-1 \approx -6\epsilon+2\eta  +2 F \frac{\Delta m^2}{m_\phi^2}\quad,\qquad
    \alpha_s \approx 4 F^2\frac{\Delta m^2}{m_\phi^2}\quad,\qquad
    \beta_s \approx 8 F^3\frac{\Delta m^2}{m_{\phi}^2}\,.
\ee
If we assume $F\sim {\cal O}(1)$, then $n_s-1$, $\alpha_s$ and $\beta_s$ are all naturally of the same order. To provide an explicit example, for $F=1/2$, $N_*=33$, and $\Delta m^2/m_\phi^2 = 0.03$, we obtain
\begin{align}\label{eq:benchmark2}
n_s=0.97\,, \quad \alpha_s= 0.03\,, \quad \text{and} \quad \beta_s = 0.03,
\end{align}
which corresponds to the purple star in Fig.~\ref{fig:running_running} and is close to our best-fit cosmology. This scenario still leaves $\Delta N = N_*-N_b  =3.5$ $e$-folds before the assumption $m_\phi > \Delta m(N)$ breaks down (corresponding in total to 10 $e$-folds of the first act of inflation). Moreover, we note that the pattern in \eqref{eq:by_proxy} is compatible with the resummed expression in \eqref{eq:spectral_ansatz_resum}.

Since we are here considering an unstable mode, depending on the precise model, it can also enhance higher-order correlation functions of the curvature perturbation, when it appears in exchange diagrams \cite{Seery:2008ax}, and leave a large non-Gaussian cosmic collider type signal \cite{Arkani-Hamed:2015bza,Lee:2016vti,Chen:2016uwp,Meerburg:2016zdz,Chen:2022vzh,Pajer:2024ckd}, which is infrared enhanced.

\subsubsection{Curvaton with proxy running}

If the observed perturbations are created by the curvaton $\sigma$, instead of the inflaton, the above formulae simplify \cite{Sloth:2014sga}. In this case the spectral tilt becomes
\be
 n_\sigma-1=-2 \epsilon+\frac{2}{3} \frac{V^{\prime \prime}}{H^2}\,,
\ee
where the $\epsilon$ comes from the background inflaton dynamics, and primes are now partial derivatives with respect to $\sigma$. If we now assume that the curvaton couples to the heavy fast-rolling field 
\be
V(\sigma, N) = \frac{1}{2}(m_\sigma^2 +\Delta m^2(N))\sigma^2\,, 
\ee
then we have
\be
n_\sigma-1=-2 \epsilon+\frac{2}{3} \frac{m_\sigma^2}{H^2}\left(1+\frac{\Delta m^2(N)}{m_\sigma^2}\right) \approx-2 \epsilon+\frac{2}{3} \frac{m_\sigma^2}{H^2}\,.
\ee
Now using that $d \log k=H d t$ and time derivatives of slow-roll parameters are order slow-roll parameters squared, we have
\be
\alpha_\sigma=\frac{d n_\sigma}{d \log k} \approx F \frac{\Delta m^2(N)}{H^2}\,,\quad \qquad  \beta_\sigma=\frac{d \alpha_\sigma}{d \log k} \approx F^2 \frac{\Delta m^2(N)}{H^2}\,,
\ee
where we assumed, for simplicity, as before $\Delta m^2 = \lambda_\chi \chi^2$, and $\chi$ is a heavy fast-rolling field with time-evolution given by (\ref{solchi}). Again if $F\sim {\cal O}(1)$, then running and running of running is of similar size.

\begin{figure}[t]
  \centering
  \begin{minipage}[c]{0.69\textwidth}
    \centering
    \includegraphics[width=\linewidth]{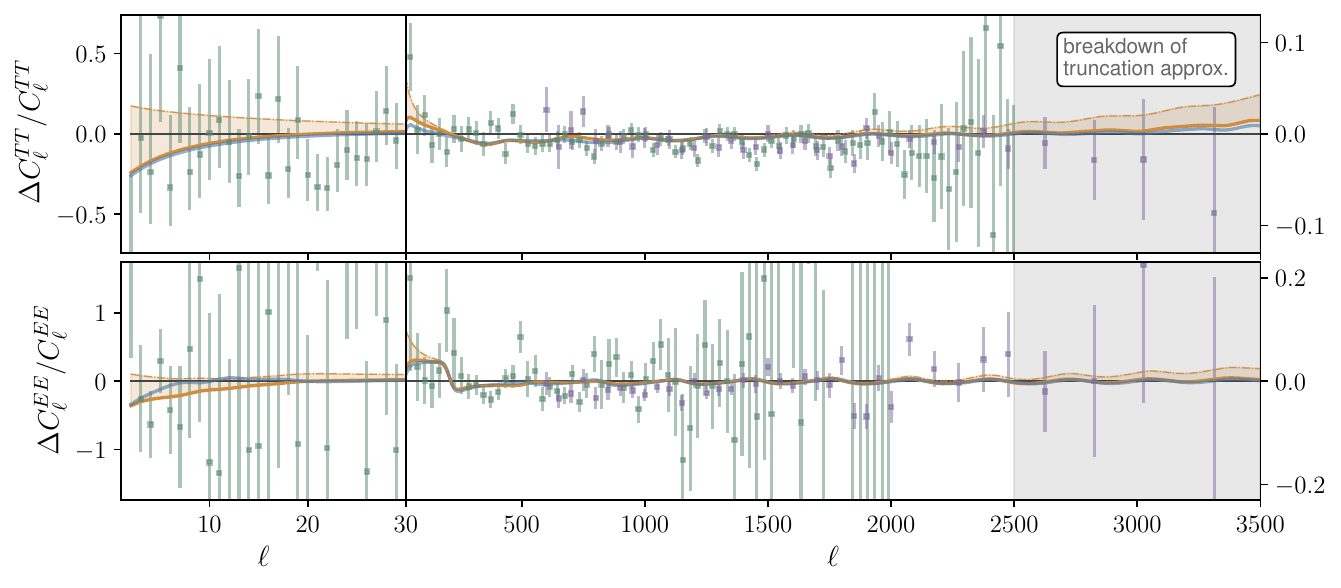}
  \end{minipage}
  \hfill
  \begin{minipage}[c]{0.3\textwidth}
    \centering
    \includegraphics[width=\linewidth]{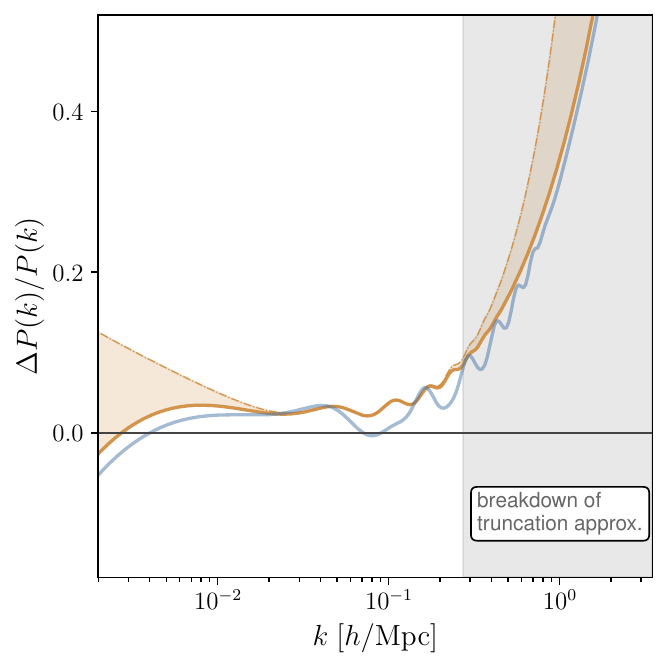}
  \end{minipage}
  \vspace{0.15cm}
  \includegraphics[width=0.9\textwidth]{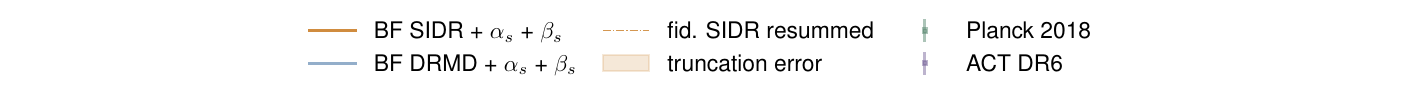}
  \caption{Best-fit SIDR (orange) and DRMD (blue) models fitted to CMB data (Planck 2018 and ACT DR6), BAO measurements (DESI DR2), and SN data (Pantheon+), shown relative to $\Lambda$CDM (black). The orange-shaded contours represent the uncertainty associated with truncating \eqref{eq:spectral_ansatz} at order $\ln^3(k/k_*)$, as estimated from the resummed expression in~\eqref{eq:spectral_ansatz_resum}. On small scales, this approximation leads to above percent-level errors in the gray-shaded region; this is also where error bars become large and the first stage of inflation could end. \textit{Left:} Temperature and polarization residuals for Planck 2018 (green) and ACT DR6 (purple). \textit{Right:} Linear matter power spectrum. The more pronounced oscillations in the DRMD case reflect the presence of DAO.
  }
  \label{fig:residuals}
\end{figure}

\section{Impact of ACT on models with extra radiation and spectral running}\label{datasec}

\subsection{Models}

We investigate to which extent cosmological models comprising a component of fluid-like dark radiation are compatible with ACT DR6 data when allowing for spectral running of the primordial adiabatic scalar perturbations.
To this end, we employ the publicly available\footnote{\href{https://github.com/NEDE-Cosmo/DRMD-CLASS}{https://github.com/NEDE-Cosmo/DRMD-CLASS}} \texttt{DRMD-CLASS} extension of the Boltzmann solver Cosmic Linear Anisotropy Solving System \texttt{CLASS}~\cite{Lesgourgues:2011re}, featuring a self-interacting dark radiation (SIDR) component produced after BBN, with the possibility of redshift-dependent momentum-drag interactions with dark matter. Our version of \texttt{DRMD-CLASS} is rebased to \texttt{CLASS} version 3.3.4 and implements spectral running as defined in \eqref{eq:spectral_ansatz}.  Using this framework, we consider three models:
\begin{enumerate}
    \item[$(i)$] the six-parameter $\Lambda$CDM model as baseline,
    \item[$(ii)$] its post-BBN SIDR extension with spectral running and running-of-running, ``SIDR + $\alpha_s$ + $\beta_s$'', and
    \item[$(iii)$] the dark radiation-matter decoupling extension of $(ii)$,  ``DRMD + $\alpha_s$ + $\beta_s$''.
\end{enumerate}

Compared to $\Lambda$CDM, model $(ii)$ introduces three additional parameters, namely $\Delta N_\mathrm{eff}$, $\alpha_s$, and $\beta_s$. Here, $\Delta N_\mathrm{eff}$ quantifies the abundance of (fluid-like) self-interacting dark radiation generated after BBN through
\begin{align}
\Delta N_\mathrm{eff} \equiv \frac{\rho_\mathrm{DR}}{\rho_{\nu,1}}\,,
\end{align}
where $\rho_\mathrm{DR}$ is the energy density of the dark radiation component and $\rho_{1\nu} = \tfrac{7}{4}\tfrac{\pi^2}{30}(T_{\nu,0}/a)^{4}$ denotes the energy density of a single Standard Model neutrino species. The dark radiation component is assumed to be tightly coupled with itself, which translates into the requirement of vanishing higher Boltzmann moments beyond its density contrast and velocity divergence. The latter two are evolved using the continuity and Euler equation with equation of state parameter and sound speed $w_\mathrm{DR} = c_\mathrm{DR}^2 = 1/3$ subject to standard adiabatic initial conditions (see \cite{Garny:2025kqj} for more details of the implementation). The parameters $\alpha_s$ and $\beta_s$ describe, respectively, the running and running of the running of the primordial scalar power spectrum, as introduced through the ansatz in Eq.~\ref{eq:spectral_ansatz}.

Model $(iii)$ further extends model $(ii)$ by including two additional parameters associated with dark-radiation--dark-matter interactions, namely the fraction of interacting dark matter $f_\mathrm{idm}$, and a characteristic redshift scale $z_\mathrm{stop}$, which controls the drag rate between dark matter and dark radiation through
\begin{align}\label{eq:drag}
\frac{\Gamma_\mathrm{drag}}{H} = c_0\, \left( 1+ \frac{\Omega_m}{\Omega_\mathrm{rad} } \frac{1}{1+z} \right)^{-1/2} \exp\left( -\frac{1+z_\mathrm{stop}}{1+z} \right)\,,
\end{align}
where $c_0$ is a constant. Within the Hot NEDE model, $c_0$ is related to the $t$-channel Compton scattering between dark radiation and dark matter fermions in the fundamental representation of a dark $SU(N)$ gauge group.
The exponential redshift-dependence arises due to a loop-induced mass splitting $\Delta m$ between charged and neutral dark matter particles, which is generated after spontaneous symmetry breaking $SU(N)\to SU(N-1)$ by a Higgs mechanism in the dark sector. Only the heavier charged particles keep interacting with dark radiation but become Boltzmann-suppressed for dark-sector temperatures $T_d$ below the mass gap, or $z < z_\mathrm{stop}\approx \Delta m /(a\, T_d) $, equivalently.

In particular, this time-dependence controls through the condition 
\be
  \Gamma_\mathrm{drag}/H\big|_{z=z_\text{dec}} = 1\,,
\ee
when the dark radiation ceases to be tightly coupled with dark matter, thereby determining the decoupling redshift $z_\mathrm{dec}$ of dark radiation and dark matter. We refer to Ref.~\cite{Garny:2025kqj} for more details on the microscopic model. On a phenomenological level, the DRMD scenario modifies the evolution of perturbations relative to the pure SIDR case, while the background evolution is identical. In particular, the DRMD model predicts a characteristic dark acoustic oscillation feature imprinted on matter perturbations~\cite{Garny:2026ish}.

\begin{figure}[t]
    \centering

    \includegraphics[width=0.985\linewidth]{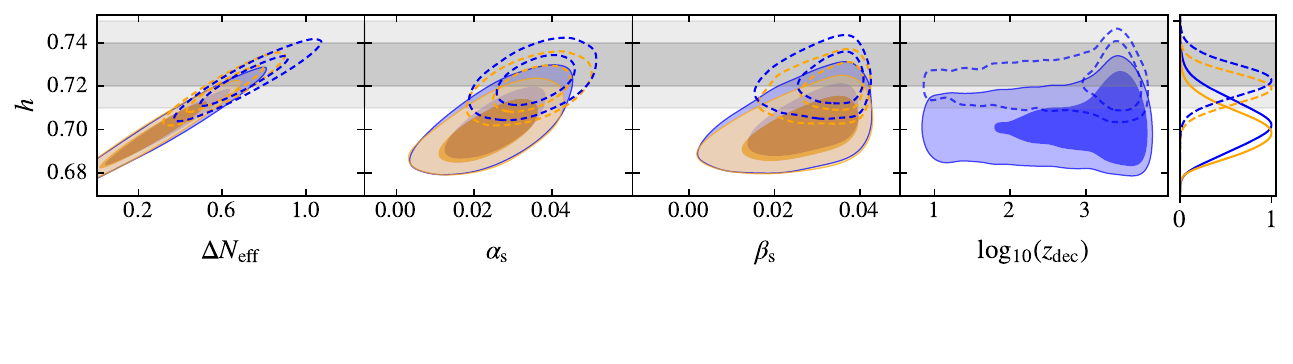}

    \vspace*{-2.5em}

    \includegraphics[width=0.985\linewidth]{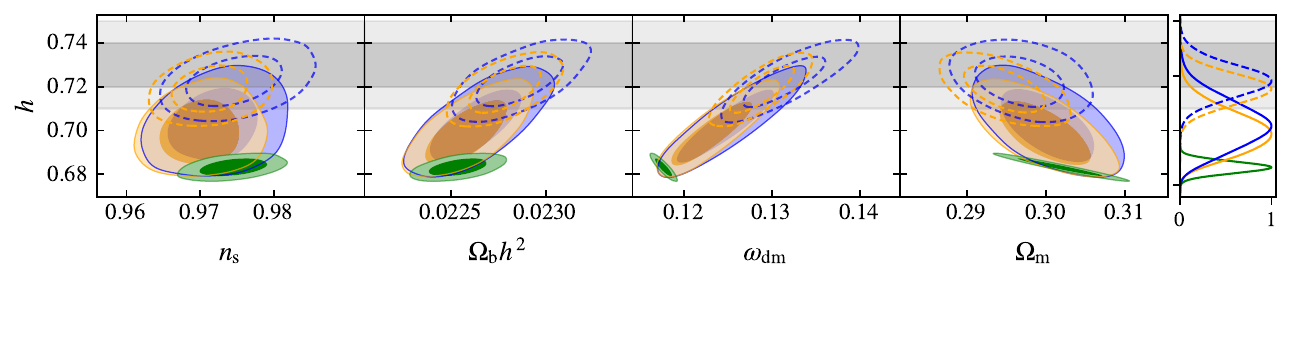}

    \vspace*{-2.5em}
\vspace{0.15cm}
    \includegraphics[width=0.55\linewidth]{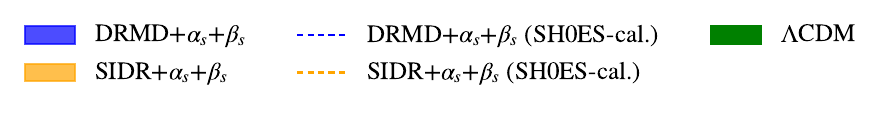}

 \vspace*{-0.5em}
    \caption{Marginalized posteriors for CMB (ACT DR6 + Planck 2018), DESI BAO (DR2), and uncalibrated Pantheon+ SNe data applied to the DRMD + $\alpha_s$ + $\beta_s$ and SIDR + $\alpha_s$ + $\beta_s$ models, including also the $\Lambda$CDM case where applicable for comparison. Vertical gray bands indicate the $68\%$ and $98\%$ CL direct $H_0$ measurement from SH$0$ES~\cite{Riess:2021jrx}, which is not used in our main analysis. For comparison, dashed contours include the SH$0$ES calibration of the SN magnitude. }
    \label{fig:rectangle}
\end{figure}

\subsection{Data and analysis}

We employ a joint Planck-ACT CMB compilation (temperature, polarization, and lensing)~\cite{Planck:2019nip,AtacamaCosmologyTelescope:2025blo}, Pantheon+ SN data~\cite{Scolnic:2021amr,Brout:2022vxf}, and DESI DR2 BAO data~\cite{DESI:2025zgx,DESI:2025zpo}. The CMB dataset combines the \textit{ACT DR6 lite} likelihood with Planck 2018 data (with cuts $\ell \leq 1000$ and  $\ell \leq 600$ for TT and (TE, EE) spectra, respectively), including the SRoll2 low-$\ell$ polarization likelihood~\cite{deBelsunce:2021mec}, and is supplemented by ACT DR6 CMB lensing data~\cite{ACT:2023kun,ACT:2023dou} (in the \textit{actplanck\_baseline} variant). We further use the Pantheon+ SN dataset in its {\it uncalibrated} form where it constrains the shape of the late expansion history for our main results, which are thus  {\it not} informed by direct $H_0$ measurements. For comparison, we also consider the calibrated form, where a prior from the SH$0$ES collaboration is imposed on the absolute SN magnitude~\cite{Riess:2021jrx} -- effectively using the distance ladder value $H_0^{\mathrm{(SH0ES)}}= 73.04 \pm 1.04\,\mathrm{km/s/Mpc}$ as a further data point. We stress again that the latter is not used in our main results, however.

For parameter inference, we use the publicly available Monte Carlo code \texttt{Cobaya}~\cite{Torrado:2020dgo}, interfaced with \texttt{DRMD-CLASS}, to sample the posterior distributions of the cosmological and model parameters.
For all three models, we sample the standard cosmological parameters
\begin{align}
\{\omega_b\,,\, \omega_\mathrm{cdm}\,,\, H_0\,,\, \ln(10^{10} A_s)\,, n_s\,,\, \tau_\mathrm{reio} \}\,,
\end{align}
with standard prior ranges supplemented by the corresponding extension parameters $\{\Delta N_\text{eff}, \alpha_s, \beta_s\}$ for model $(ii)$ and $\{\Delta N_\text{eff}, f_\text{idm}, z_\text{stop}, \alpha_s, \beta_s\}$ for model $(iii)$ as introduced above. For the extension parameters we apply the following flat priors (if applicable)
\begin{equation}
\Delta N_\mathrm{eff} \in [0,3]\,, \qquad 
f_\mathrm{idm} \in [0,1]\,, \qquad 
\log_{10}(z_\mathrm{stop}) \in [2,5]\,, \qquad 
\alpha_s,\, \beta_s \in [-0.1,0.1]\,.
\end{equation}
We take one of the three neutrinos to be massive with $m_3 = 0.06\,\mathrm{eV}$ and $T_3 = 0.716\,T_\gamma$. In the case of DRMD, we choose an initial drag rate ${\mathcal{G}/(aH)|_\mathrm{ini} = 10^7 }$, where $\mathcal{G}/a =  \Gamma_\mathrm{drag}$ as defined in \eqref{eq:drag}. This choice, which is degenerate with $z_\mathrm{stop}$~\cite{Garny:2025kqj}, ensures a tightly coupled fluid at initial (post-BBN) times. 

Best-fit values are obtained by running the profile likelihood code \texttt{PROSPECT}~\cite{Holm:2023uwa} interfaced with \texttt{Cobaya}. Moreover, we quantify the residual Hubble tension from the distribution of $\Delta \equiv H_0^{\mathrm{(MCMC)}} - H_0^{\mathrm{(SH0ES)}}$. To that end, we calculate $ P(\Delta \le 0)$ and convert it into the corresponding Gaussian-equivalent tension, quoted in units of $\sigma$. In the same way we determine the evidence for $\alpha_s,\beta_s \geq 0$ based on their 1D posteriors.

\subsection{Results}

We find that for both the DRMD + $\alpha_s$ + $\beta_s$ and SIDR + $\alpha_s$ + $\beta_s$ models, the combination of ACT DR6 + Planck 2018 CMB, DESI DR2 BAO and uncalibrated Pantheon+ SN data indicates a preference for spectral running with positive $\alpha_s$ and $\beta_s$ parameters, with significance of $2.9\sigma$ for $\alpha_s>0$ and $2.6\sigma$ for $\beta_s>0$. This is accompanied with largely relaxed constraints for the post-BBN value of fluid-like $\Delta N_\text{eff}$,
\bea\label{eq:result}
  \Delta N_\text{eff} &=& 0.30^{+0.14}_{-0.18}\ \  (68\% \,\text{CL})\,\, \text{and} \,\, \Delta N_\text{eff}< 0.576\,(95\%\,\mathrm{CL}) \,\, \text{for} \,\, \text{SIDR}+\alpha_s+\beta_s\,,\\
  \Delta N_\text{eff} &=& 0.37^{+0.18}_{-0.18}\ \ (68\% \,\text{CL})\,\, \text{and}\,\, \Delta N_\text{eff} < 0.682\,(95\%\,\mathrm{CL}) \,\,\text{for} \,\,\text{DRMD}+\alpha_s+\beta_s\,,\nn
\eea
where results for all other model parameters are collected in Tab.~\ref{tab:MCMC}. In Fig.~\ref{fig:running_running}, we show for the SIDR model that $\Delta N_\mathrm{eff}$ is positively correlated with spectral running. The largest running allowed by the data reaches $\alpha_s \sim \beta_s \sim 0.04$ and is associated with $\Delta N_\mathrm{eff} \sim 0.5$; similar behaviour is found for DRMD. By contrast, in the absence of running, $\Delta N_\mathrm{eff}$ is driven to zero, in agreement with the literature~\cite{AtacamaCosmologyTelescope:2025nti}.

The marginalized 2D posteriors for both models are shown in Fig.~\ref{fig:rectangle}, displaying the correlation of new model parameters (upper panel) as well as various $\Lambda$CDM parameters (lower panel) with the dimensionless Hubble constant $h$. As expected, the regions in parameter space with non-zero $\Delta N_\text{eff}$ lead to larger inferred values of $h$, allowing for a reduction or even resolution of the tension with direct SH$0$ES measurements that exists at high significance within $\Lambda$CDM. In particular, for SIDR + $\alpha_s$ + $\beta_s$ this tension is reduced to $2.2\sigma$. For DRMD + $\alpha_s$ + $\beta_s$, SH$0$ES data are compatible with ACT DR6 + Planck 2018 CMB, DESI DR2 BAO and uncalibrated Pantheon+ SNe at a level below $2\sigma$. We therefore display also a result from jointly combining all of these data sets by the dashed contours in Fig.~\ref{fig:rectangle}, showing a strong preference for extra radiation combined with spectral running in that case, as expected. Notably, this is achieved with adding only three extra parameters to the $\Lambda$CDM model for SIDR + $\alpha_s$ + $\beta_s$, while two more parameters enter for DRMD + $\alpha_s$ + $\beta_s$.

The underlying reason for the success of these models in accommodating large values of $\Delta N_\mathrm{eff}$ is illustrated in Fig.~\ref{fig:residuals}, which depicts the best-fit residuals for SIDR (orange) and DRMD (blue), both including spectral running and shown relative to $\Lambda$CDM. Without spectral running, fluid-like radiation suppresses power on small scales (see the residual plot in~\cite{Garny:2025kqj}). A positive running of the spectral tilt offsets this suppression up to multipoles of order $\ell \sim 3000$, such that both the SIDR and DRMD models produce residuals of similar size to those of $\Lambda$CDM. Indeed, our $\chi^2$ analysis indicates an overall improvement of about $\Delta \chi^2 \approx -6$ for the combined data sets, strengthening to $\Delta \chi^2 \approx -30$ upon inclusion of the SH$0$ES prior.

\begin{figure}[t]
  \centering
  \includegraphics[width=0.49\textwidth]{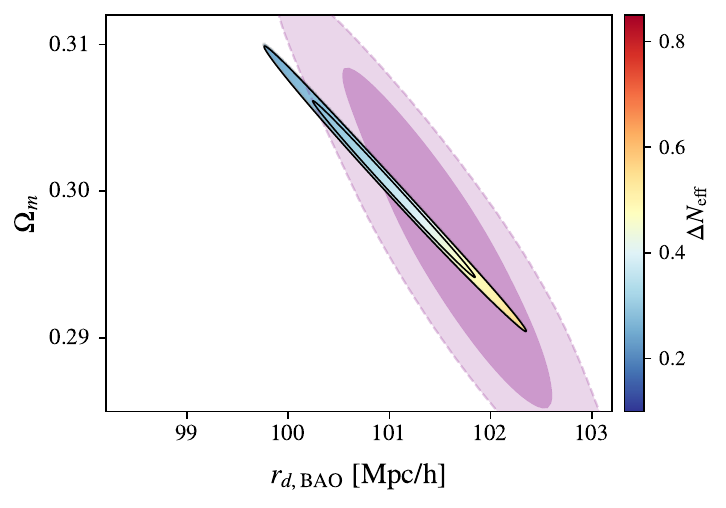}\hfill
    \includegraphics[width=0.49\textwidth]{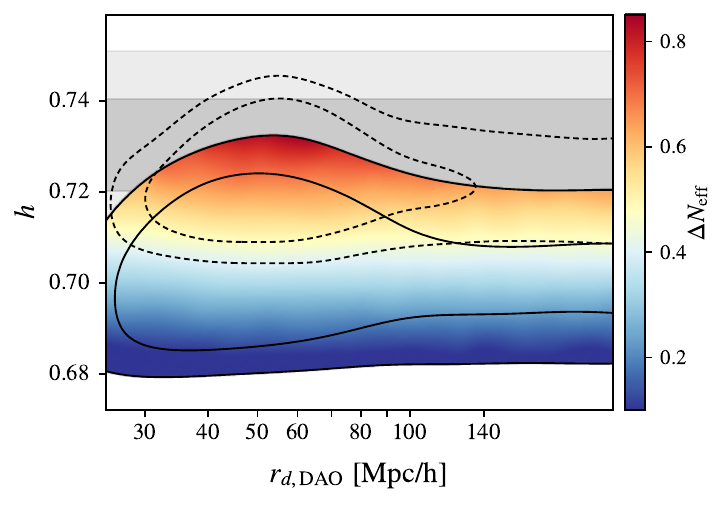}\hfill
\vspace{0.15cm}
      \includegraphics[width=0.55\linewidth]{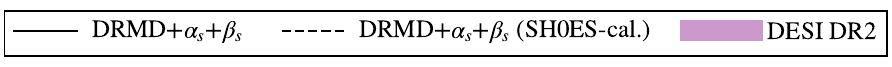}

  \caption{Marginalized 2D posteriors in the DRMD + $\alpha_s$ + $\beta_s$ model for CMB (ACT DR6 + Planck 2018), BAO (DESI DR2), and uncalibrated SN (Pantheon+) data. Dashed contours include the SH$0$ES calibration of the SN magnitude. \textit{Left:} Larger values of $\Delta N_\mathrm{eff}$ lead to smaller $\Omega_m$ in agreement with the late-time DESI DR2 constraint (purple). \textit{Right:} Within DRMD, DAO around $r_{d,\mathrm{DAO}} \sim 60\, \mathrm{Mpc}/h$ enable larger values of $H_0$. The gray-shaded band corresponds to the SH$0$ES measurement of~$H_0$.}
  \label{fig:twoplots}
\end{figure}

It is instructive to consider also the parameter space spanned by the BAO drag horizon scale $r_{d,\text{BAO}}$ in units relative to the Hubble length $1/h$ as well as $\Omega_m$, to which BAO measurements are primarily sensitive. The left panel of Fig.~\ref{fig:twoplots} shows the DESI DR2 result in this parameter plane, as well as the region preferred by the DRMD + $\alpha_s$ + $\beta_s$ model. Both are well compatible, with best agreement for larger values of $\Delta N_\text{eff}$. This is not the case within $\Lambda$CDM, for which CMB data prefer smaller values of $r_{d,\text{BAO}}$ and larger values of $\Omega_m$ than DESI DR2 BAO data. While we did not strictly quantify the extent to which this CMB-BAO tension is relaxed, our results suggest that they are well compatible within DRMD + $\alpha_s$ + $\beta_s$. We find a very similar behaviour in SIDR + $\alpha_s$ + $\beta_s$, in line with previous findings regarding the interplay of the $H_0$ with the BAO-CMB tension~\cite{Poulin:2025nfb,Garny:2025szk,Garny:2026ish,Jhaveri:2026bla,AtacamaCosmologyTelescope:2025nti,SPT-3G:2025vyw}.

A particular prediction of the DRMD model is a DAO feature due to the decoupling of dark matter--dark radiation interactions around matter-radiation equality. This characteristic prediction has been previously identified based on the data considered in this work, except for ACT DR6, and without considering spectral running \cite{Garny:2026ish,Garny:2025szk}. We find that, when including ACT DR6 as well as spectral running, this feature persists, with a similar preferred range for the DAO drag horizon scale $r_{d,\text{DAO}}\sim 60$\,Mpc$/h$, as shown in the right panel of Fig.~\ref{fig:twoplots}. Since photons do not participate in DAO, the latter are primarily imprinted in matter perturbations (as well as the largely unobservable dark radiation perturbations), while leaving only a relatively mild impact on the CMB (see also the more detailed discussion in~\cite{Garny:2026ish}). Moreover, the CMB is affected by DAO at scales relevant around the DRMD decoupling redshift $z_\mathrm{dec}$, which is preferred to occur close to matter-radiation equality (upper right panel of Fig.~\ref{fig:rectangle}). Those angular scales $\ell\ll 10^3$ are probed primarily by Planck CMB data, making it plausible that adding ACT DR6 has only an indirect, mild impact on the predicted DAO. Instead, the DAO could be tested based on full-shape DESI data, beyond the scope of this work. In this context, one may wonder about the interplay with possible imprints in the matter power spectrum related to the inflationary dynamics responsible for generating the required amount of spectral running. However, the scales at which potential features from the transition from the first to the second stage of inflation are expected are beyond those relevant for BAO measurements, and better probed by smaller scales such as Lyman-$\alpha$ and future CMB lensing observations. We indicate the earliest possible onset of this transition by the gray band in Fig.~\ref{fig:residuals}, corresponding to the regime where the uncertainties from truncating higher-order running terms in \eqref{eq:spectral_ansatz} become non-negligible (orange-shaded). Where exactly this transition occurs will depend on the precise model and describing it would require a more detailed ansatz for the primordial power spectrum, taking the resummed form in \eqref{eq:spectral_ansatz_resum} (dash-dotted) in an intermediate regime  before accounting for the perturbative breakdown at even smaller scales. We note however that for ACT data employed in our current analysis the error bars become large for $\ell \gtrsim 2500$, exceeding the theoretical uncertainty in magnitude and making our simple modelling applicable. On the other hand, we do not include the recent compressed Lyman-$\alpha$ constraint from~\cite{Chaves-Montero:2026hqd}. Besides not being directly applicable to the dark-radiation extensions of $\Lambda$CDM considered here, it would also lie within the gray band in the right panel of Fig.~\ref{fig:residuals}.

\section{Conclusion}\label{concl}

The ACT collaboration has interpreted the DR6 data release as showing no evidence for new light degrees of freedom at the same time as confirming the Hubble tension. We have shown that these conclusions are highly model dependent. 

In single field slow-roll inflation, the running of the spectral index with comoving scales is higher order in slow-roll, and the spectral index can therefore be approximated as effectively scale-independent. However, from a theory perspective, it has long been argued that single-field slow-roll inflation is unlikely to proceed uninterrupted for $60$ $e$-folds, and as a consequence a scale-dependent primordial spectral index is more natural than not, since the interruptions of slow-roll inflation will be associated with a sharp increase in the spectral index at scales corresponding to modes exiting the horizon close to the interruptions.  We have  briefly discussed how the sharp increase in the spectral index can show up as large running and running of the running in models of inflation where a first act of inflation ends by a resonant particle production of gauge fields or by a tachyonic instability. 

Allowing therefore for a scale-dependent spectral index, we show that the ACT data becomes compatible with the presence of new light degrees of freedom as large as\footnote{Such a large value of $\Delta N_{\textrm{eff}}$ is consistent with BBN bounds, if the dark radiation is created from the latent heat of a phase-transition after BBN as in \cite{Garny:2024ums, Garny:2025kqj}.} $\Delta N_{\textrm{eff}} \sim 0.6$ in the form of self-interacting dark radiation (see Tab.~\ref{tab:MCMC} and \eqref{eq:result} for detailed results). This result was obtained in a three-parameter extension of $\Lambda$CDM, allowing for a component of (fluid-like) SIDR produced after BBN as well as running, $\alpha_s$, and running of the running, $\beta_s$, of the primordial spectral index $n_s$, when fitted to ACT DR6 + Planck 2018 CMB data, DESI DR2 BAO data, and Pantheon+ uncalibrated SN data. At the same time, we find evidence of non-vanishing running, $\alpha_s>0$, and running of the running, $\beta_s>0$, with $2.9\sigma$ and $2.6\sigma$ significance respectively.

Turning to the Hubble tension, we find that in this three-parameter extension of $\Lambda$CDM, the Hubble tension is reduced to $2.2\sigma$, almost resolving the Hubble tension. Adding a component of interacting dark matter, which interacts with the self-interacting dark radiation, only to undergo dark radiation-matter decoupling (DRMD) close to matter equality \cite{Garny:2025kqj}, the Hubble tension is fully resolved to below $2\sigma$. The DRMD model also predicts the existence of dark acoustic oscillations with a dark drag horizon $r_{\textrm{d,DAO}}\approx 60$ Mpc/h. Our earlier studies already found hints for dark acoustic oscillations on this scale \cite{Garny:2025szk,Garny:2026ish}.

Since the models of SIDR and DRMD with running and running of the running of the spectral index are not showing significant tensions with the SH$0$ES data, we also consider the combination of ACT DR6 + Planck 2018 CMB, DESI DR2 BAO and Pantheon+ SN data with SH$0$ES within these models. Once we include the SH$0$ES data, the evidence for running becomes very strong, with a value that is nominally offset from zero by more than ten times the width of the 68\% CL interval. This is accompanied by a strong $\sim 5 \sigma$ evidence for $\Delta N_\mathrm{eff}>0$. From a statistical point of view, this may be argued to justify adding three additional parameters to the $\Lambda$CDM model. Of course, this is a very strong conclusion, which one should carefully stress-test. A first worry is that the result, driven by the small-scale ACT data, could also be hinting at unknown systematics at small scales, where astrophysical foregrounds become important, although the ACT collaboration has performed a highly sophisticated analysis in testing for residual foreground contaminations of their cosmological signal. 

Without SH$0$ES data, the evidence for running (running of running) is still $2.9\sigma$ $(2.6\sigma)$, so to fully mitigate the evidence beyond $\Lambda$CDM, one would need to find new previously unknown systematics significantly contaminating the ACT and the SH$0$ES data. In addition, if one were to insist on $\Lambda$CDM, DESI results would also have to be explained by unknown systematics. Interestingly, the DESI anomaly could be explained simultaneously to SH$0$ES within the DRMD model due to the impact of the DAO feature on the extraction of the BAO scale~\cite{Garny:2025szk,Garny:2026ish}, with the required DAO scale matching the one predicted from requiring a solution of the Hubble tension within this model. Somewhat more speculatively, we also point out that the origin of spectral running due to particle production from gauge fields could be accommodated within the gauged dark sector underlying DRMD when complementing it by an axion field. Interestingly, if the latter couples also to the visible sector, this theoretical setup could even be connected to a further anomaly that has been reported from CMB observations, including ACT DR6 data, related to a cross correlation of $E$ and $B$ polarization modes that could be explained by birefringence~\cite{Diego-Palazuelos:2025dmh}.
Thus, taken together, the anomalies in the ACT, DESI and SH$0$ES data, may be interpreted as coherently pointing at new physics, like DRMD, beyond $\Lambda$CDM. In the future, data from the Simons Observatory~\cite{SimonsObservatory:2018koc} will be instrumental to test this scenario, along with large-scale structure probes related to DAOs as well as further possible imprints of inflationary dynamics proceeding in several stages on smaller scales, such as those probed by Lyman-$\alpha$ observations.

\subsection*{Acknowledgements}

MG acknowledges support by the Excellence Cluster ORIGINS, which is funded by the Deutsche Forschungsgemeinschaft (DFG, German Research Foundation) under Germany’s Excellence Strategy – EXC-2094 - 390783311 as well as the DFG Collaborative Research Centre ``Neutrinos and Dark Matter in Astro- and Particle Physics'' (SFB 1258). FN was supported by VR Starting Grant 2022-03160 of the Swedish Research Council. MSS acknowledges Nordita for their kind hospitality through the Nordita corresponding fellow program.

\newpage

\begin{appendix}
\clearpage
\section*{Summary of MCMC results}
\begin{table}[h]
\centering
\fontsize{8}{9.5}\selectfont
\label{tab:MCMC}
\renewcommand\cellset{\renewcommand\arraystretch{1.1}}
\begin{tabular}{c|c|c|c|c|c}  

\rule{0pt}{3ex}&  \multicolumn{1}{c|}{$\Lambda$\textbf{CDM}}   & \multicolumn{2}{c|}{\textbf{SIDR}+ $\alpha_s$+$\beta_s$}  &  \multicolumn{2}{c}{\textbf{DRMD}+$\alpha_s$+$\beta_s$}   \\ 
\hline
\hline

\rule{0pt}{4ex}&Base&Base&\makecell{Base\\ + SH${0}$ES-cal.}&Base&\makecell{Base\\ + SH${0}$ES-cal.}\\
\hline

\rule{0pt}{4ex} \# parameters & 6  & 6+3 & 6+3 & 6+5 & 6+5\\ 
\hline
\rule{0pt}{3ex}\makecell{$\omega_b$} & \makecell{$0.02254\pm 0.00010$\\$(0.02253)$} & \makecell{$0.02259\pm 0.00014$\\$(0.02260)$} & \makecell{$0.02278\pm 0.00013$\\$(0.02284)$} & \makecell{$0.02263^{+0.00015}_{-0.00017}$\\$(0.02268)$} & \makecell{$0.02286\pm 0.00015$\\$(0.02296)$}
\\
\hline
\rule{0pt}{3ex}\makecell{$\omega_{\mathrm{cdm}}$} & \makecell{$0.11764\pm 0.00063$\\$(0.11764)$} & \makecell{$0.1233^{+0.0026}_{-0.0034}$\\$(0.1242)$} & \makecell{$0.1288\pm 0.0024$\\$(0.1301)$} & \makecell{$0.1250^{+0.0033}_{-0.0040}$\\$(0.1266)$} & \makecell{$0.1315^{+0.0028}_{-0.0034}$\\$(0.1330)$}
\\
\hline
\rule{0pt}{3ex}\makecell{$H_0\,\mathrm{[km/sec/Mpc]}$} & \makecell{$68.31\pm 0.26$\\$(68.31)$} & \makecell{$69.94^{+0.84}_{-1.1}$\\$(70.18)$} & \makecell{$71.88\pm 0.70$\\$(72.46)$} & \makecell{$70.3^{+1.0}_{-1.2}$\\$(70.8)$} & \makecell{$72.27\pm 0.76$\\$(72.76)$}
\\
\hline
\rule{0pt}{3ex}\makecell{$\ln(10^{10} A_s)$} & \makecell{$3.061\pm 0.011$\\$(3.061)$} & \makecell{$3.051^{+0.012}_{-0.013}$\\$(3.050)$} & \makecell{$3.043^{+0.012}_{-0.013}$\\$(3.036)$} & \makecell{$3.052^{+0.012}_{-0.014}$\\$(3.048)$} & \makecell{$3.045^{+0.012}_{-0.014}$\\$(3.047)$}
\\
\hline
\rule{0pt}{3ex}\makecell{$n_\mathrm{s}$} & \makecell{$0.9744\pm 0.0030$\\$(0.9741)$} & \makecell{$0.9701\pm 0.0037$\\$(0.9693)$} & \makecell{$0.9714\pm 0.0034$\\$(0.9707)$} & \makecell{$0.9721\pm 0.0042$\\$(0.9698)$} & \makecell{$0.9748^{+0.0040}_{-0.0046}$\\$(0.9752)$}
\\
\hline
\rule{0pt}{3ex}\makecell{$\tau_{\mathrm{reio}}$} & \makecell{$0.0635^{+0.0057}_{-0.0066}$\\$(0.0623)$} & \makecell{$0.0640^{+0.0059}_{-0.0067}$\\$(0.0636)$} & \makecell{$0.0637^{+0.0057}_{-0.0068}$\\$(0.0615)$} & \makecell{$0.0644^{+0.0058}_{-0.0069}$\\$(0.0636)$} & \makecell{$0.0638^{+0.0057}_{-0.0069}$\\$(0.0653)$}
\\
\hline
\rule{0pt}{3ex}\makecell{$\Delta N_\mathrm{eff}$} & -- & \makecell{$0.30^{+0.14}_{-0.18}\,(68\%\,\mathrm{C.I.})$\\$< 0.576\,(95\%\,\mathrm{C.I.})$\\$(0.35)$} & \makecell{$0.61\pm 0.12$\\$(0.69)$} & \makecell{$0.37\pm 0.18\,(68\%\,\mathrm{C.I.})$\\$< 0.682\,(95\%\,\mathrm{C.I.})$\\$(0.46)$} & \makecell{$0.71\pm 0.14$\\$(0.79)$}
\\
\hline
\rule{0pt}{3ex}\makecell{$\alpha_s$} & -- & \makecell{$0.0242\pm 0.0082$\\$(0.0274)$} & \makecell{$0.0346^{+0.0068}_{-0.0062}$\\$(0.0408)$} & \makecell{$0.0258^{+0.0091}_{-0.0080}$\\$(0.0317)$} & \makecell{$0.0357^{+0.0069}_{-0.0063}$\\$(0.0397)$}
\\
\hline
\rule{0pt}{3ex}\makecell{$\beta_s$} & -- & \makecell{$0.0264^{+0.012}_{-0.0061}$\\$(0.0306)$} & \makecell{$0.0330^{+0.0070}_{-0.0026}$\\$(0.0397)$} & \makecell{$0.0268^{+0.012}_{-0.0057}$\\$(0.0353)$} & \makecell{$0.0327^{+0.0071}_{-0.0026}$\\$(0.0396)$}
\\
\hline
\rule{0pt}{3ex}\makecell{$\log_{10}(z_\mathrm{dec})$} & -- & -- & -- & \makecell{$2.91^{+0.88}_{-0.22}$} & \makecell{$\left[3.17,\,3.67\right]$}
\\
\hline
\rule{0pt}{3ex}\makecell{$f_\mathrm{idm}$} & -- & -- & -- & \makecell{$< 0.0227$\\$(0.0068)$} & \makecell{$< 0.0337$\\$(0.0191)$}
\\
\hline
\hline
\rule{0pt}{3ex}\makecell{$\Omega_m$} & \makecell{$0.3018\pm 0.0035$\\$(0.3004)$} & \makecell{$0.2996\pm 0.0039$\\$(0.2981)$} & \makecell{$0.2947\pm 0.0034$\\$(0.2913)$} & \makecell{$0.3002\pm 0.0040$\\$(0.2982)$} & \makecell{$0.2968\pm 0.0037$\\$(0.2946)$}
\\
\hline
\rule{0pt}{3ex}\makecell{$r_{d,\mathrm{BAO}}\, \mathrm{[Mpc/h]}$} & \makecell{$100.80\pm 0.47$} & \makecell{$101.10\pm 0.52$} & \makecell{$101.75\pm 0.46$} & \makecell{$101.04\pm 0.53$} & \makecell{$101.49\pm 0.50$}
\\
\hline
\rule{0pt}{3ex}\makecell{$r_{d,\mathrm{DAO}}\, \mathrm{[Mpc/h]}$} & -- & -- & -- & \makecell{$\left[13.252,\,187.226\right]$} & \makecell{$\left[30.117,\,82.087\right]$}
\\
\hline
\rule{0pt}{3ex}\makecell{$\Delta\chi^2\;(\mathrm{wrt}\;\Lambda\mathrm{CDM})$} & \makecell{$0.00$} & \makecell{$-6.07$} & \makecell{$-29.63$} & \makecell{$-6.30$} & \makecell{$-32.58$}\\
\hline
\rule{0pt}{4ex}{$H_0$ tension (Bayesian) }&$4.5\sigma$  & 2.2$\sigma$ &--& 1.8$\sigma$ &--\\ 
\end{tabular}
	\caption{\footnotesize Summary of cosmological parameter constraints for $\Lambda$CDM, SIDR + $\alpha_s$ + $\beta_s$, and DRMD + $\alpha_s$ + $\beta_s$. ``Base'' refers to the dataset combination consisting of CMB (ACT DR6 + Planck 2018), SN (uncalibrated Pantheon+), and BAO (DESI DR2) data. Two-sided constraints are quoted at 68\% confidence, while one-sided limits are given at 95\% confidence. When extended tails shift the posterior mean outside the 68\% interval, we instead report the median-centered credible interval.
    }
\end{table}

\end{appendix}

\bibliography{ref}
\end{document}